\documentclass[useAMS,usenatbib]{mn2e}

\usepackage{graphicx}

\usepackage{deluxetable}

\newcommand{\ltappeq}{\raisebox{-0.6ex}{$\,\stackrel{\raisebox{-.2ex}{$\textstyle <$}}{\sim}\,$}}
\newcommand{\approxgt}{\mathrel{\spose{\lower 3pt\hbox{$\sim$}} \raise 2.0pt\hbox{$>$}}}

\newcommand{\mags}{{\rm\thinspace mag}}

\title[IPHAS Catalogue of H$\alpha$ Emitters]{The IPHAS Catalogue of 
H$\alpha$ Emission Line Sources in the Northern Galactic Plane}

\author[A.R. Witham et al.]{
A. R. Witham,$^1$
C. Knigge,$^1$
J. E. Drew,$^2$
R. Greimel,$^{3,4}$
D. Steeghs$^{5,6}$
\newauthor
B. T. G\"ansicke,$^5$
P. J. Groot,$^7$
A. Mampaso$^8$
\\
$^{1}$ School of Physics \& Astronomy, University of Southampton, Highfield, SO17 1BJ, U.K \\
$^{2}$ Imperial College London, Blackett Laboratory, Exhibition Road, London, SW7 2AZ, U.K \\
$^{3}$ Isaac Newton Group of Telescopes, Apartado de correos 321, E-38700 Santa Cruz de la Palma, Spain \\
$^{4}$ Institute of Physics, University of Graz, Universit\"atsplatz 5, 8010 Graz, Austria\\
$^{5}$ Department of Physics, University of Warwick, Coventry CV4 7AL, U.K. \\
$^{6}$ Harvard-Smithsonian Center for Astrophysics, 60 Garden Street, Cambridge, MA 02138, USA \\
$^{7}$ Department of Astrophysics/IMAPP, Radboud University Nijmegen, P.O. Box 9010, 6500 GL, Nijmegen, The Netherlands  \\
$^{8}$ Instituto de Astrof\'{i}sica de Canarias, 38200 La Laguna, Tenerife, Spain \\
\\
}

\begin{document}

\date{2005 August 5}

\pagerange{\pageref{firstpage}--\pageref{lastpage}} \pubyear{2005}

\maketitle

\label{firstpage}

\begin{abstract}

We present a catalogue of point-source H$\alpha$\ emission line
objects selected from the INT/WFC Photometric H$\alpha$ Survey of the
Northern Galactic Plane (IPHAS). The catalogue covers the magnitude
range $13 \leq r^\prime \leq 19.5$\ and includes northern
hemisphere sources in the Galactic latitude range $-5^\circ < b <
5^\circ$. It is derived from $\sim 1500$~deg$^2$ worth of imaging
data, which represents 80\% of the final IPHAS survey area. The
electronic version of the catalogue will be updated once the full
survey data becomes available. In total, the present catalogue contains 4853 point sources that
exhibit strong photometric evidence for H$\alpha$ emission. We have so far
analyzed spectra for $\sim300$ of these sources, confirming more than
95\% of them as genuine emission-line stars. A wide range of stellar
populations are represented in the catalogue, including early-type
emission line stars, active late-type stars, interacting binaries,
young stellar objects and compact nebulae.

The spatial distribution of catalogue objects shows overdensities near
sites of recent or current star formation, as well as possible 
evidence for the warp of the Galactic plane. Photometrically, the   
incidence of H$\alpha$ emission is bimodally distributed in
$(r^\prime-i^\prime)$. The blue peak is made up mostly of early-type
emission line stars, whereas the red peak may signal an increasing
contribution from other objects, such as young/active low-mass stars.
We have cross-matched our H$\alpha$-excess catalogue against the
emission-line star catalogue of \citet{1999A&AS..134..255K}, as well
as against sources in SIMBAD. We find that fewer than 10\% of our
sources can be matched to known objects of any type. Thus IPHAS is
uncovering an order of magnitude more faint ($r^\prime > 13$)
emission line objects than were previously known in the Milky Way.

\end{abstract}


\section{Introduction}

Large-scale H$\alpha$ imaging surveys have traditionally focused on
extended emission line sources, such as the nebulosities associated
with areas of intense star formation. However, many interesting
classes of stars also display H$\alpha$ emission and, in principle,
can therefore be efficiently identified in such surveys as
point-source H$\alpha$-excess objects. More often than not, line
emission is associated with stars in very early or very late
evolutionary stages that are short-lived and relatively poorly
understood. Line emission is also often associated with binaries
experiencing mass transfer and accretion. The list of line-emitting
classes of stars includes post-asymptotic and some asymptotic giant
branch (AGB) stars, compact planetary nebulae, luminous blue
variables, hypergiants, 
Wolf-Rayet stars, classical Be stars, active late-type dwarfs,
interacting binaries and a wide range of young stellar objects. Most
of these object samples are inhomogeneous, and some contain very few
identified members. This scarcity and heterogeneity places obstacles
in the way of testing current models for the formation and evolution
of these systems. Large-scale H$\alpha$ surveys can be used to expand
the known samples and to improve their homogeneity. Surveys of the
highly populous Galactic Plane should be particularly successful in
this regard.

Most previous photographic H$\alpha$ surveys suffered from bright
limiting magnitudes and/or small survey areas. Examples of such
surveys include the H$\alpha$ observations of the Large and Small Magellanic
Clouds by
\citet{1976MmRAS..81...89D} and the Kyoto photographic survey of the
northern Milky Way (\citealt{1982aanm.book.....U}). Recent CCD imaging
surveys such as the Virginia Tech H-alpha and [S\,II] Imaging Survey
of the Northern Sky (\citealt{1998PASA...15..147D,
  1999AAS...195.5309D}) and the Southern H$\alpha$ Sky Survey Atlas
(\citealt{2001PASP..113.1326G}) cover large areas of the sky, but at
relatively low spatial resolution. The Wisconsin H$\alpha$ Mapper
Northern Sky Survey (\citealt{2003ApJS..149..405H}) has
spectroscopically surveyed the Galaxy with a resolution of
$1^\circ$. This has enabled the distribution and kinematics of diffuse
H\,II to be mapped, but point source emitters could not be
investigated. 

This leaves an obvious gap for deep Galactic H$\alpha$ surveys
covering large areas at high spatial resolution.  This gap has
been partially filled by the 
photographic AAO/UKST H$\alpha$ Southern Galactic Plane Survey (Parker
et al 2005), which covers the latitude range, $|b| \leq 10^{\rm o}$,
down to $\sim$19.5\mags\ at a spatial resolution of 1\arcsec-2\arcsec.
The INT/WFC Photometric H$\alpha$ Survey (IPHAS, 
\citealt{2005MNRAS.362..753D}) is
a highly analogous survey of the northern Galactic Plane, in terms of
both depth and spatial resolution. Now close to completion, its
coverage in broadband $r'$, $i'$ and narrowband H$\alpha$ is limited
to the Galactic latitude range $|b| \leq 5^{\rm o}$.  It differs also
in being digital, allowing greatly superior photometric calibration,
and the opportunity to photometrically select emission line stars
across the northern Plane with high confidence. Here we present the
first catalogue of IPHAS emission line stars, timed to accompany the 
first major release to the world community of general IPHAS photometry
(Gonz\'{a}lez-Solares et al. 2008, in preparation).

The most extensive previous catalogue of emission-line stars in the
northern Galactic plane has been compiled by \citet[hereafter 
KW99]{1999A&AS..134..255K}. The KW99 catalogue contains 4174 objects
within the latitude range $|b| \leq 10^o$. Eighty per cent of the
objects fall in the latitude range $|b|\ltappeq5^o$. The catalogue is
expected to be complete to $\sim13$\mags, although 27 per cent of the
objects are fainter than this, and 13 per cent do not have a measured
brightness. Down to $\sim$ 13th magnitude, early-type emission line
stars dominate the lists compiled by KW99, accounting for around
three-quarters of all objects listed.  

The catalogue presented here covers the magnitude range $13 \leq r'
\leq 19.5$ mag.  It is easily cross-matched to source lists in other
wavebands and is already providing the basis for a range of
spectroscopic follow-up programmes. In Section~2, we briefly describe
the IPHAS photometric observing 
strategy. Section~3 discusses the selection algorithm used to
construct our catalogue. Section~4 provides an overview of the
catalogue and of the
spatial distribution of the catalogue objects. In Section~5, we 
analyze the magnitude and colour distributions of our H$\alpha$-excess
sources. Section~6 discusses the relative contributions of different
stellar types to the catalogue, based on matches to the KW99 catalogue
and to SIMBAD objects, and comments on preliminary results from our
own spectroscopic follow-up observations. In Section~7, we summarize
our main results and conclusions.  

\section{Observations}

\label{obs}

The data used to construct our point-source H$\alpha$-excess catalogue
are composed of the IPHAS Galactic plane observations obtained between
September 2003 and January 2007. After rejecting observations failing
the IPHAS data quality criteria (see Section 3.1), these data cover
about 80 per cent of the final survey area. Once the survey has been
completed, we will update the online version of the catalogue to cover
the final area. 

Full details of the IPHAS observing strategy, data reduction,
calibration and matched catalogue generation may be found in
\citet{2005MNRAS.362..753D}; \citet{2001NewAR..45..105I} and
\citet{1985MNRAS.214..575I}. Briefly, all observations were obtained
using the Wide Field Camera (WFC) on the  Isaac Newton Telescope
(INT), which gives a spatial pixel size of $0.33$\arcsec x
$0.33$\arcsec\ over a field of view of approximately 0.3 square
degrees. Photometry was carried out using a set of three filters,
comprising a narrowband H$\alpha$\ filter and the broadband Sloan
$r^\prime$\ and $i^\prime$\ filters. Exposure times were 120\,s for
the H$\alpha$\ images, 10\,s for the $i^\prime$ band images, and 30\,s
[10\,s] seconds for all $r^\prime$ band images obtained after [before]
June 2004.  

Note that in all of our analysis below, we only work with sources
brighter than $r^\prime = 19.5$. In principle, IPHAS goes deeper
than this, as the average $r^\prime$ band $3\sigma$ magnitude limit in
the IPHAS fields used to construct the catalogue is
$^\prime = 21\pm1$. However, the increased photometric scatter and
incompleteness at the faintest magnitudes makes it counter-productive 
to push our H$\alpha$-excess selection to this depth. We also adopt a
bright limit of $r^\prime = 13$\ to exclude saturated objects.

The tiling pattern of the IPHAS survey is designed to ensure that
objects are not routinely lost in the chip gaps of the INT/WFC
CCDs. This is achieved by taking two observations for each pointing;
here, we refer to these as the direct field and the offset field. The
offset fields are located 5\,arcmin W and 5\,arcmin S of the direct
field centres, and are identified by the letter ``o'' following the
numerical field code. The existence of the offset fields means that
the vast majority of objects in the survey ($\sim 95\%$) are 
observed at least twice. As explained further below, we exploit this
to improve the robustness of our H$\alpha$-excess selection.

\section{Selection Strategy}

\label{sec_ss}

\subsection{Data Quality Constraints}

Only data from fields that pass strict quality acceptance criteria were  
used to construct our catalogue. The two most important criteria are: 

\begin{enumerate}

\item the seeing in each band must not be greater than 2\,arcsec;

\item the average stellar ellipticity in the images of each band must
  not be greater than 0.2.    

\end{enumerate} 

This leaves 12959 IPHAS fields, providing an effective sky coverage of 
$\sim1500\,\mathrm{deg}^2$. We also apply additional quality cuts at
the level of individual sources. Thus each object in our catalogue

\begin{enumerate}
\item  must be detected at least twice in the survey
  (usually in a direct field and its offset field); 
\item must be selected as an H$\alpha$-excess source in 
  {\em both} detections (see Section~\ref{select} for details on the
  selection algorithm); 
\item must have an $r^\prime$ magnitude in the range $13
  < r^\prime < 19.5$;
\item must not be flagged as severely blended; 
\item must be classified as stellar in the $i^\prime$ band,
  and stellar or probably stellar in the $r^\prime$ and H$\alpha$
  bands.  
\end{enumerate}

Cut (v) is imposed because our catalogue is intended mainly as a
list of {\em stellar} emission line sources. Objects classified as
probably stellar in the $r^\prime$ and H$\alpha$ bands are
nevertheless included to allow for compact nebulae and emission line
stars with marginally resolved H$\alpha$ shells (such as novae). 

As a final precaution, we assign a flag 'c' to all objects in our
catalogue for which the two $r^\prime$ measurements, the two
$r^\prime-i^\prime$ colours 
or the two $r^\prime$-H$\alpha$ colours disagree by more than 0.3
mag. This could signal genuine variability, but can also occur if the
photometric calibration of one field is incorrect due to
partial cloud cover (a few such fields may be expected to survive all
our quality cuts). Our selection algorithm is actually fairly
insensitive to photometric zero-point offsets, but the photometry of
emitters  with inconsistent magnitudes/colours should be 
regarded with caution. All of the analysis in this paper is restricted
to stars and emitters whose magnitudes and colours agree to better
than 0.3 mag, unless explicitly stated otherwise in the relevant
figure captions.

Only one set of photometric data is listed for each object in the
catalogue, as derived from the highest quality field in which the
object was detected. Throughout this paper, we will use the term field
to refer to a single direct or offset field, and pointing to refer to
them jointly (e.g. an object detected in a pointing is detected in 
{\em both} a field and its offset field).

\subsection{Selecting H$\alpha$ Emitters}

\label{select}

\begin{figure*}

\includegraphics[angle=-90,width=\textwidth]{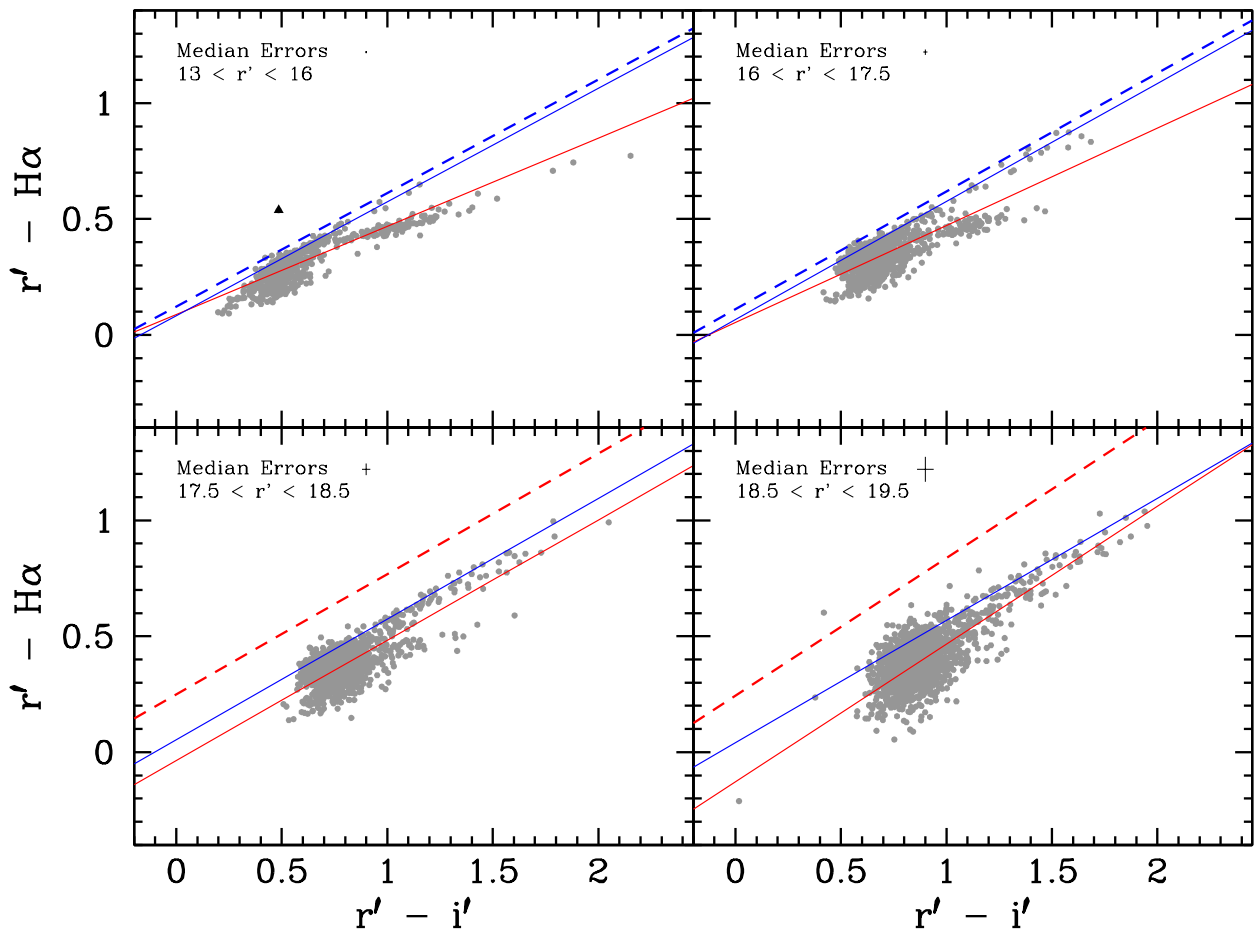}

\caption{\label{fig_selection} An illustration of the selection
criteria used to identify strong emission line objects via
colour-colour plots. The data shown here are all from the IPHAS
field 2373. The data are split up into four magnitude bins, as shown
in the four panels. The median errors of the datapoints in each bin
are shown by the black crosses near the top of each
plot. Objects with H$\alpha$\ excess should be located near the top
of the colour-colour plots.  The thin red lines illustrate the
original least squares fit to all the data (grey points). The thin blue
lines represent the final fits to the upper locus of points obtained
by applying an iterative $\sigma$-clipping technique to the initial
fit.  The actual cuts used to select H$\alpha$ emitters are shown by the
thick dashed lines. If the cut was based on the initial [final] fit, it is
shown in red [blue]. Objects selected as H$\alpha$ emitters must be
located above the cut and are shown as large triangles.  Note that the
cut lines shown here are only approximate, as the actual selection
criterion also considers the errors on each individual
datapoint. This explains, for example, why an object in the bottom
right panel is not selected despite clearly lying above the cut
line.}

\end{figure*}

H$\alpha$-excess sources are selected from the 12959 IPHAS fields using the
algorithm described in \citet{2006MNRAS.369..581W}. The selection
process is illustrated in Fig.\,\ref{fig_selection}. Briefly, we generate
$(r^\prime-i^\prime)$~versus~$(r^\prime - \rmn{H}\alpha)$
colour-colour plots for each magnitude bin in each field. We then
carry out an initial straight-line least-squares fit to all objects in
each magnitude bin. However, many IPHAS fields exhibit (at least) two
separate stellar loci in the colour-colour plane, due to differential
reddening and/or contributions from both main-sequence stars and
giants. We therefore use an iterative $\sigma$-clipping technique to
force the fit onto the uppermost locus of points in the colour-colour
diagram; this upper locus generally represents the unreddened main
sequence. In cases where the final fit is poorer than the initial fit
(e.g. in fields containing only a single stellar locus), we revert
back to the initial fit. Once the appropriate fit for each magnitude
bin has been found, we identify objects significantly above the fit as
likely H$\alpha$\ emitters. In doing so, we take into account both the
scatter of points in the stellar loci and the errors on the colours of
each individual datapoint.  

As explained in \citet{2006MNRAS.369..581W}, this automated selection
process works well most of the time, but it can occasionally fail. The
usual cause is a failure of the algorithm to satisfactorily identify
the true upper edge of the stellar locus in the colour-colour
diagram. In the present context, the most serious problem arising from
this is that several of the objects belonging to this locus may be
selected erroneously as H$\alpha$-excess sources. In the worst of
these cases, the number of objects selected from a field will be
abnormally high. 

In practice, we find that the probability of algorithm failure becomes 
significant for fields where 10 or more objects were selected
automatically. We therefore visually inspect the colour-colour plots
for all such fields and replace the automated selections with manually
identified ones. In doing so, we also keep track of fields where the
colour-colour plots look genuinely unusual, so that it is hard to
identify true emitters with confidence; objects selected from such
fields are flagged in the catalogue. In total, 328 fields have been
manually inspected, and 231 of the objects selected by the automated
algorithm in these fields have been culled. However, the final
catalogue still contains 752 manually-selected objects from these fields.

In spite of all the data quality controls we impose and the care we
have taken to design a robust selection procedure, it 
is, of course, impossible to guarantee that every object in our
catalogue is a definite H$\alpha$ emitter. However, as discussed in
Section~\ref{spec}, spectroscopic follow-up observations of $\sim300$
objects from our catalogue with $r^\prime \leq 18$\ have confirmed
the presence of H$\alpha$ emission in about 97 per cent of this sample. 

It is much harder to establish the completeness of our selection
procedure, i.e. the fraction of H$\alpha$\ emitters selected as a
function of magnitude, broad-band colour and equivalent
width. Qualitatively, it is clear that the minimum equivalent width (EW) above which an
object will be selected increases with increasing magnitude and
reddening (see Fig.\,\ref{fig_selection} and discussion in
\citet{2005MNRAS.362..753D}). However, the only way
to determine the completeness quantitatively is to obtain spectra for 
a large, representative sample of objects (not just likely
emitters). Such a large-scale completeness analysis has not yet been
carried out. The more limited follow-up observations we have analyzed
so far indicate that, as expected, our selection method is robust and
conservative, rather than highly complete.

\section{Results}

The full catalogue of H$\alpha$ emitters is only available in
electronic form and can be obtained at http://www.iphas.org or via
VizieR at http://vizier.u-strasbg.fr/viz-bin/VizieR. As noted
in Section~\ref{obs}, it currently covers 80\% of the final IPHAS
survey area, but will be updated once the survey has been
completed. The first few entries in the catalogue are given in
Table~\ref{cat_short}.

\begin{table*}

\scriptsize

\begin{minipage}{\linewidth}

\begin{center}




\caption{\label{cat_short}IPHAS positions, magnitudes and colours of
the first few objects in the catalogue of point-source H$\alpha$
emitters. The full catalogue is available from http://www.iphas.org or
from VizieR}

\begin{tabular}{llllll}


\hline \\[-2ex]

   \multicolumn{1}{c}{\textbf{IPHAS name/position}} &

   \multicolumn{3}{c}{\textbf{IPHAS photometry}} &

   \multicolumn{1}{c}{\textbf{flag}\tablenotemark{a}} &


   \multicolumn{1}{c}{\textbf{SIMBAD match flag}\tablenotemark{b}} \\ 


   \multicolumn{1}{c}{J[RA(2000)+Dec(2000)]} &

   \multicolumn{1}{c}{$r^\prime$} &

   \multicolumn{1}{c}{$r^\prime-i^\prime$} &

   \multicolumn{1}{c}{$r^\prime-$H$\alpha$} &

   \multicolumn{1}{c}{} &

   \multicolumn{1}{c}{} \\

  \hline

   \\[-1.8ex]


J000000.18+645440.7 & $17.231\pm0.005$ & $ 0.861\pm0.009$ & $ 0.680\pm0.009$ &     & 										      \\

J000039.05+623316.6 & $14.168\pm0.002$ & $ 0.217\pm0.003$ & $ 0.271\pm0.003$ & c   & 										      \\

J000137.14+620423.4 & $16.944\pm0.005$ & $ 0.410\pm0.011$ & $ 0.361\pm0.010$ &     & 										      \\

J000213.37+645424.6 & $16.969\pm0.010$ & $ 0.856\pm0.013$ & $ 1.346\pm0.011$ & ck  & C/Em$^\ast$=RI$^\ast$ ($^\ast$,V$^\ast$) 								      \\

J000220.16+594538.7 & $14.447\pm0.001$ & $ 0.439\pm0.002$ & $ 0.427\pm0.002$ &     & 										      \\

J000250.83+634633.7 & $15.388\pm0.002$ & $ 0.760\pm0.004$ & $ 0.651\pm0.004$ &     & 										      \\

J000259.61+620916.3 & $14.983\pm0.002$ & $ 0.723\pm0.004$ & $ 0.794\pm0.004$ &     & 										      \\

J000301.71+621020.3 & $14.276\pm0.001$ & $ 0.725\pm0.002$ & $ 0.764\pm0.002$ & k   & UC/Em$^\ast$ 									      \\

J000315.54+625423.0 & $15.257\pm0.002$ & $ 0.519\pm0.004$ & $ 0.475\pm0.004$ &     & 										      \\

J000335.07+632946.2 & $13.792\pm0.002$ & $ 0.566\pm0.003$ & $ 0.675\pm0.002$ & k   & UC/Em$^\ast$ 									      \\

...\\

\hline

\footnotetext[1]{This column is used to flag any special circumstances
related to the photometry or if the object is in the vicinity of a
KW99 object. Possible entries in this column are as follows: c - the
object does not have consistent magnitudes in any of the IPHAS
fields used to construct the catalogue.  This can be as a result of
source variability or poor photometric calibration as a result of
observing in less than ideal conditions. m - the object is located
in a field where manual selection of emitters took place. o - the
object is located in a field where manual selection of emitters took
place and additionally the distribution of objects in the
colour-colour plots from this field looked unusual. k - the object
has been matched with a previously known H$\alpha$ emitter in KW99
with a matching radius of up to 30 arcsec.} 

\footnotetext[2]{An entry in this column indicates that the object is
matched with a SIMBAD object.  Entries begin with either UC/ or C/
followed by the short-hand main SIMBAD classification.  UC/
designates objects whose main object type falls within the 8 broad
classifications defined in Section~\ref{simb_match}. The maximum
matching radius for these objects is listed in
Table~\ref{main_match}. C/ designates those objects whose main
object type is not one of the 8 main types. These objects have a
generally finer classification and have a maximum matching radius of
10\,arcsec. For the C/ objects the other possible object types
listed in Simbad are also given in parenthesis. The C/Em$^\ast$
entries are the SIMBAD objects whose main object type is not
Em$^\ast$ but have Em$^\ast$ listed as another object type.  If
these objects are within 10\,arcsec of the IPHAS source then the
object is designated by C/Em$^\ast$= followed by the main object
type.  If the object is located at a distance greater than
10\,arcsec (but still within the 45\,arcsec matching radius used for
objects whose main type is Em$^\ast$), then the object is designated
by C/Em$^\ast\sim$ followed by the main object type.}

\end{tabular}									             

\end{center}

\end{minipage}

\normalsize

\end{table*}

\subsection{Basic Number Statistics}

Our H$\alpha$-excess catalogue contains 4853 objects selected from
12959 fields. This represents an order of magnitude improvement in the
number of faint ($r^\prime \geq 13$) H$\alpha$ emitters in the Northern
Galactic Plane (Section~\ref{nature}). The final catalogue constructed 
from the completed IPHAS survey should therefore contain at least 6000
objects, even without any improvements to the selection technique. It
is nevertheless worth working towards such improvements: extrapolation
of the KW99 and \citet{1971PW&SO...1a...1S} surveys suggest that
10,000 - 40,000 objects could be uncovered by IPHAS if a more relaxed 
(or simply better) selection technique could be used.

The number of objects in our catalogue corresponds to an average
surface density of $\sim3$ H$\alpha$\ emitters per square degree in
the Northern Galactic Plane. Similarly, we find that roughly 1 in 7000
stars are selected as H$\alpha$ emitters. The conservative nature
of our selection technique means that these numbers should be viewed
as lower limits. 

\subsection{The Spatial Distribution of H$\alpha$ Emitters}

\label{spatial}

\subsubsection{The 2-D Distribution of H$\alpha$ Emitters}

\label{spatial_2D}

\begin{figure*}

\centering

\includegraphics[angle=0,width=1.0\linewidth]{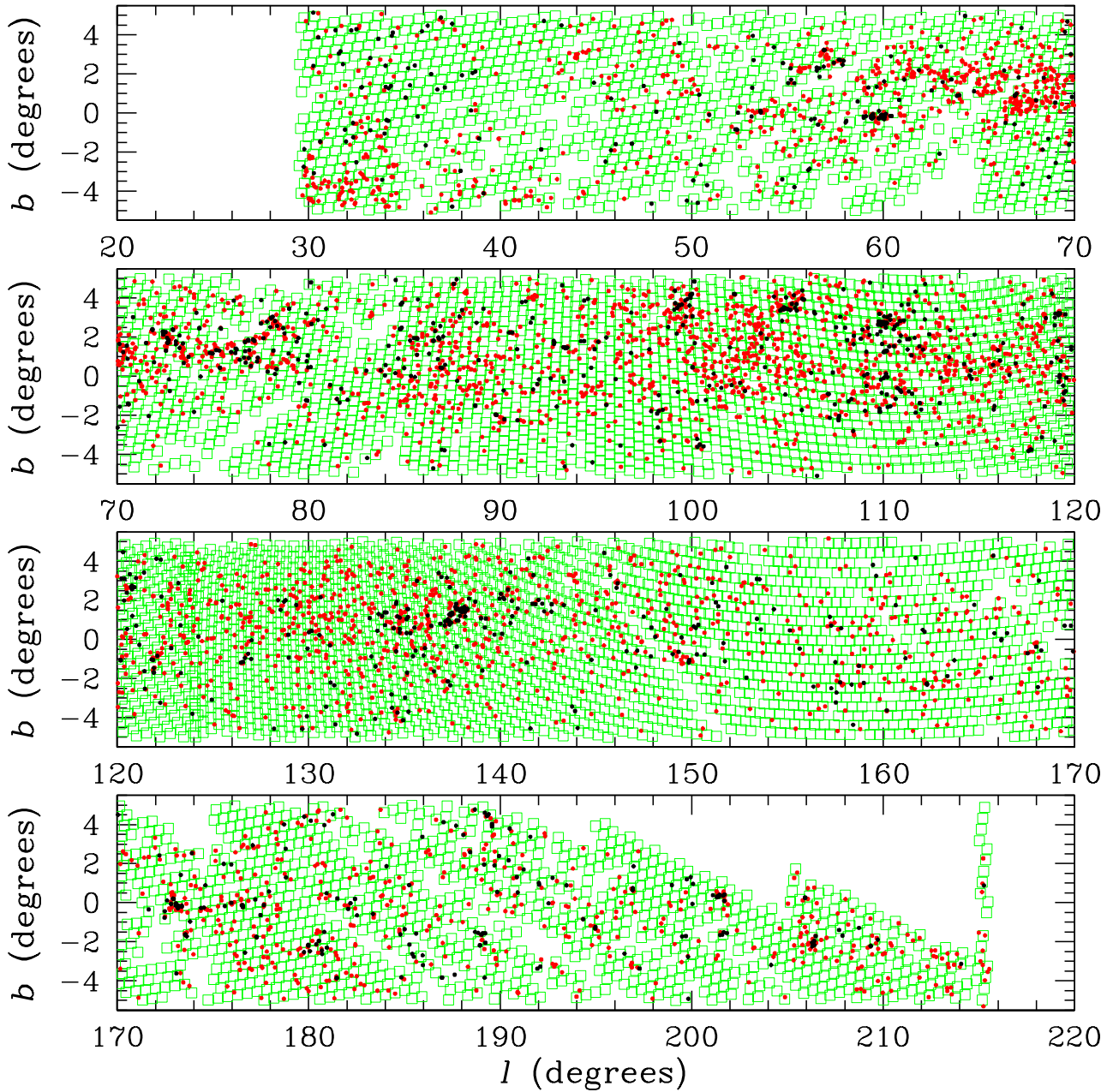}


\caption{The distribution of H$\alpha$ emitters in Galactic longitude
and latitude. The emitters are shown as red points if brighter than
$r^\prime = 18$, and black points if fainter. The IPHAS direct
fields are shown by green squares (offset fields are not shown). All
emitters are shown here, including those with flagged with 'c' in Table~1.}

\label{fig:gal_4_panel}

\end{figure*}

Fig.\,\ref{fig:gal_4_panel} shows the 2-D distribution of all 
catalogue objects in Galactic latitude and longitude, along with the 
distribution of all parent fields from which the catalogue was
created. On average, the surface density of H$\alpha$-excess objects
is highest near the Galactic Equator, particularly in the longitude
range 60--$140^\circ$. This reflects the increasing density of
H$\alpha$ emitters near the mid-plane of the Galactic disc. The
surface density of emitters is also rather non-uniform (cf KW99). This
is to be expected, since the density of H$\alpha$ emitters will be
enhanced near clusters, OB associations, etc. 

We have visually identified some of the most prominent overdensities
in Fig.\,\ref{fig:gal_4_panel}, concentrating especially on areas rich
in faint emitters. These regions are as follows: 

\begin{enumerate}

\item $l\sim60$, $b\sim0$: This region coincides with the OB
  association Vul OB1 and, on smaller scales, the central cluster of
  this association. This cluster, NGC 6823, is already known to
  harbour at least 2 Be stars \citep{1999A&AS..136..313S}. 

\item $l\sim68$, $b\sim1$: This overdensity of mainly bright emitters
  is close to the H{\sc ii} region Sharpless 2-98, which is ionized by
  the Wolf-Rayet star W130 \citep{cich}. The WR star itself is in the KW99
  catalogue, as are several other objects in the vicinity of this
  star-forming complex. 

\item $l\sim99.5$, $b\sim3.5$: This overdensity is coincident with the
  young open cluster Trumpler 37, which belongs to the Cep OB2
  \citep{1990AJ.....99.1536M}. Several H$\alpha$ emitters have already 
  been found in this region
  (e.g. \citealt{1990Ap&SS.174...13K}). 

\item $l\sim105$, $b\sim4.0$: This overdensity appears to be connected to
  the LDN 1188 dark cloud complex. This may in turn belong to the
  Cepheus bubble and is known to contain several T Tauri stars and
  other H$\alpha$ emitters
  \citep{1995A&A...300..525A}. \citet{2005MNRAS.362..753D} present
  multi-object spectroscopy entered on $l\sim105.6$, 
  $b\sim4.0$ and find 29 definite H$\alpha$ emitters in the
  magnitude range $17\leq r^\prime \leq20.5$. 

\item $l\sim110.5$, $b\sim3.0$: This area lies within the Cep OB3
  association which contains many previously known H$\alpha$
  emitters (see, for example, \citealt{2001Ap&SS.275..441M}). 

\item $l\sim135$, $b\sim1.5$: This location lies close to several BRCs
  that are known to harbour H$\alpha$\ emitters. Two examples are BRC 5 and
  7 \citep{2002AJ....123.2597O}. The OB association Cas OB6 also lies
  nearby. The central cluster of this association, IC 1805, is known
  to contain early-type emission-line stars
  \citet{1999A&AS..136..313S}. 

\item $l\sim138$, $b\sim1.5$: This region coincides with bright rim
  cloud (BRC) 14, which is already known to harbour 47 H$\alpha$ emitters
  \citep{2002AJ....123.2597O}. 

\item $l\sim173$, $b\sim-0.25$: The OB association Aur OB2 is located
  in this region of the sky, along with the H\,II region IC
  417. However, we have not been able to identify clusters, BRCs or
  other object of smaller spatial scale near this location.  

\end{enumerate}

\begin{figure*}

\centering

\includegraphics[angle=0,width=1.0\linewidth]{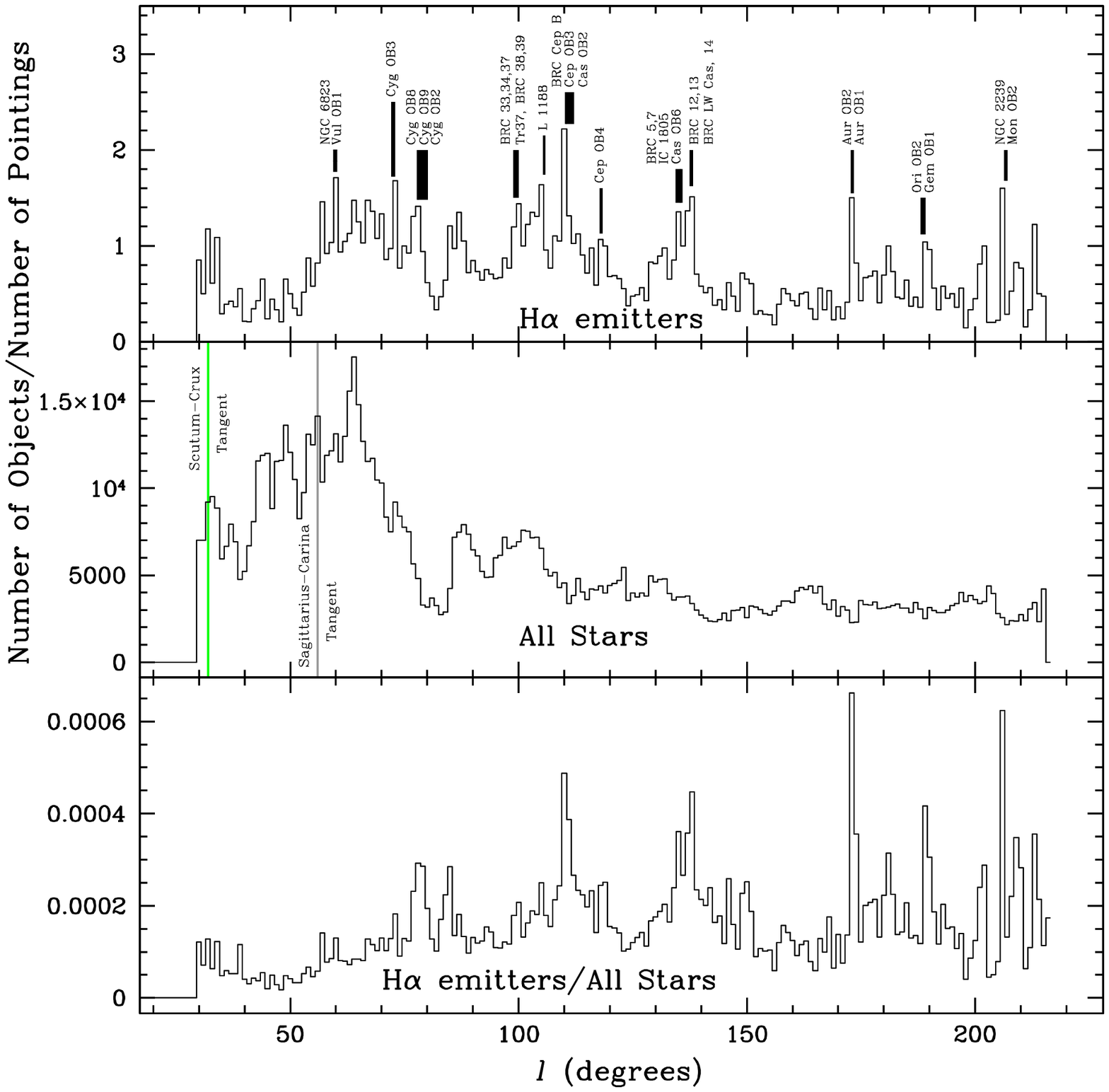}


\caption{The distribution of H$\alpha$ emitters in Galactic
  longitude. The top panel shows the distribution of H$\alpha$
  emitters divided by the number of pointings in each spatial bin. OB
  associations, BRCs, and clusters whose position coincide with peaks
  in the distribution are labelled. The width of the lines marking the
  peaks illustrates the spread in the central locations of the
  different regions that are appropriate to that peak. The middle
  panel shows the distribution of all stars divided by the
  number of pointings in each spatial bin.  The bottom panel shows the
  ratio of the two distributions. The bin width for all the
  distributions is $1^\circ$.  The green and grey vertical lines in
  the middle panel show the location of the tangents to the
  Scutum-Crux and Sagittarius-Carina arms respectively
  (\citealt{1992ASSL..180..131B}, \citealt{2003A&A...397..133R}).}

\label{fig:gal_lon}

\end{figure*}

\begin{figure}

\centering

\includegraphics[angle=0,width=1.0\linewidth]{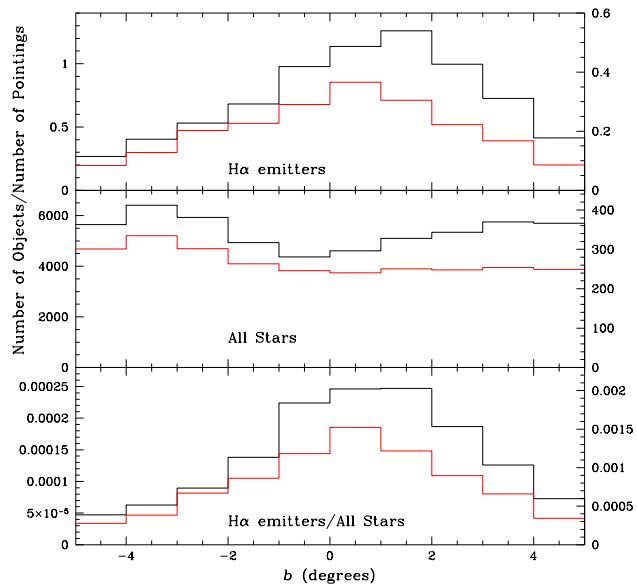}


\caption{The distribution of H$\alpha$ emitters in Galactic
  latitude. The black histograms (corresponding to the left vertical
  scales) show the distributions of objects irrespective of $r^\prime$
  band  magnitude. The red histograms (corresponding to the right
  vertical scales) show the distribution of the bright objects only
  ($r^\prime \leq15$). The top panels show the distributions of H$\alpha$
  emitters (divided by the number of pointings in each spatial bin). 
  The middle panels show the corresponding distributions of all
  stars. The bottom panels show the ratios of emitter and all-star
  distributions.}

\label{fig:gal_lat}

\end{figure}

Other spatial features traced by the catalogue of H$\alpha$ emitters
can be more readily identified by collapsing the information in
Fig.\,\ref{fig:gal_4_panel} to one dimension. The histograms in
Fig.\,\ref{fig:gal_lon} and Fig.\,\ref{fig:gal_lat} show the
distribution of objects in Galactic longitude and latitude. What is
actually plotted in these figures is the number of H$\alpha$\ emitters
per longitude/latitude bin, divided by the number of IPHAS pointings
in that bin. This normalization helps to remove the effects of spatial
variations in survey coverage. 

\subsubsection{The Longitude Distribution of H$\alpha$ Emitters}

The top panel of Fig.\,\ref{fig:gal_lon} shows the distribution of
catalogue objects in Galactic longitude. This distribution shows a
multitude of peaks and troughs over the region surveyed. OB
associations, clusters, bright rimmed clouds, and other regions of
active star formation can cause local peaks in the
distribution. Several of the most prominent peaks have been labelled
with the names of the objects that are likely responsible for these
overdensities. 

The middle panel shows the distribution of all stars, again
divided by the number of catalogue pointings in each spatial bin. The 
green and grey vertical lines at $\ell = 33^{\rm o}$ and $\ell = 55^{\rm o}$ in
this panel mark the locations of the tangent directions reported by
\citet{1992ASSL..180..131B}, that are now seen as associated with 
the Scutum-Crux and Sagittarius-Carina arms (see
e.g. \citealt{2003A&A...397..133R}).  In the top panel of
Fig.\,\ref{fig:gal_lon}, there is evidence of a peak in the range
$30^{\rm o} < \ell < 35^{\rm o}$ in the surface density of H$\alpha$
emitters (see also Fig.\,\ref{fig:gal_4_panel}). The all-star sample
also peaks here, and so both emission line and normal stars appear to
be somewhat more abundant in the vicinity of the first tangent point.

In between the $\ell = 33^{\rm o}$ and $\ell = 55^{\rm o}$ tangents,
there are relatively few candidate emission line stars, even though
the surface density of normal stars climbs. This is where the
Sagittarius-Carina arm might be expected to be most prominent. The
scarcity of emission line stars in this region may be due to 
a combination of high reddening and limited H$\alpha$ emission
equivalent width (under 30 to 40 \AA , figure 6 in
\citealt{2005MNRAS.362..753D}). The highly extincted Aquila Rift lies
within this region, for example. Weak emission line stars within or
beyond such local absorbing structures could be well concealed. This
will warrant further investigation. 

At longitudes beyond $\sim60^{\rm o}$, the surface densities of both
emission line stars and normal stars show a general decline,
interrupted by much finer scale structure. Over this range, out to
$\ell \simeq 215^{\rm o}$, the major Galactic structures are expected
to be the Perseus arm and the rather more uncertain Outer, or
Norma-Cygnus, arm. The most prominent localised feature in the
all-star sample distribution is the broad and deep dip at $\ell \sim
80^{\rm o}$ linked with the highly obscured Cygnus-X region. This drop is
less marked in the H$\alpha$ emitter distribution, and hence appears
as a relative (if doubled) peak in the ratio distribution. Cygnus X is
known to harbour much star formation (e.g. \citealt{1993ApJ...405..706O}). 

\subsubsection{The Latitude Distribution of H$\alpha$ Emitters}

The top panel of Fig.\,\ref{fig:gal_lat} shows the distribution of
catalogue objects in Galactic latitude. The black histogram reveals
that the density of H$\alpha$-excess objects actually peaks at
latitudes between $1^\circ$ and $2^\circ$. This contrasts with
the result from KW99, who find that the distribution of bright
H$\alpha$\ emitters peaks close to the Galactic Equator. The
difference is mainly due to the inclusion of fainter H$\alpha$
emitters in the IPHAS catalogue. When we consider only 
objects with $r^\prime \leq 15$\ (red histograms in
Fig.\,\ref{fig:gal_lat}), we find a similar latitude distribution to
KW99. The central dip in the latitude distribution of the all-star
samples is due to the high extinction near the Galactic equator.

\begin{figure}

\centering

\includegraphics[angle=270,width=1.0\linewidth]{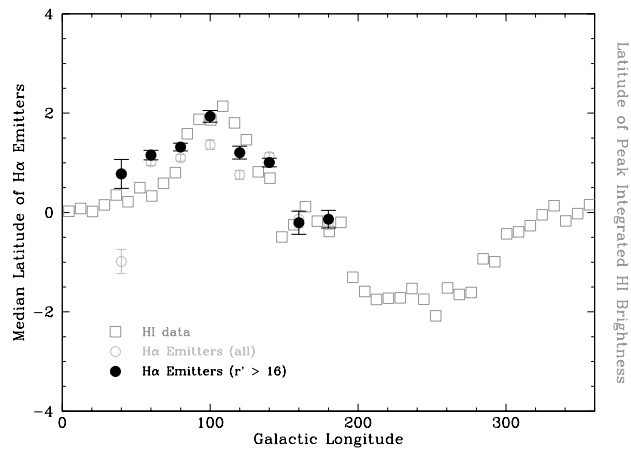}

\caption{The median latitude as a function of longitude for all
H$\alpha$\ emitters (open circles) and faint H$\alpha$\ emitters (
($r^\prime > 16$; solid circles), compared to the peak latitude
of integrated H~{\sc i} brightness (open squares; measured from Fig. 3 in
\protect\citealt{1994ApJ...429L..69F}). The signature of the Galactic
warp is obvious in the H~{\sc i} data, and probably also be present in 
the H$\alpha$ data.}

\label{fig:warp_peak}

\end{figure}

The latitude distribution of H$\alpha$\ emitters is also longitude
dependent. This is illustrated in Fig.\,\ref{fig:warp_peak}, which
shows the median latitude of our H$\alpha$\ emitters as measured in
20$^\circ$\ longitude bins. These data can be compared directly to the
latitude of peak integrated H~{\sc i} brightness of the Galactic disk
(open squares in Fig.\,\ref{fig:warp_peak}; taken from
\citealt{1994ApJ...429L..69F}). The H~{\sc i} data clearly show the
signature of the Galactic warp.  The strong similarity between the
H~{\sc i} peak brightness and the median H$\alpha$ emitter latitude
seen in Fig.\,\ref{fig:warp_peak} suggests the latter may also trace the
warp. This is quite reasonable, given that our catalogue is probably
dominated by young objects (and particularly early-type stars; see
Sections~\ref{bimodal} and \ref{spec}). Both the integrated infra-red
light of the stellar Galactic disk \citep{1994ApJ...429L..69F}, as
well as OB stars specifically \citep{1991Ap&SS.177..399M}, are
already known to trace the warp.

Fig.\,\ref{fig:warp_peak} also suggests that the probable warp
signature is somewhat clearer if only fainter emitters ($r^\prime >
16$) are considered. This, too, is plausible. As shown in
\citet{2006A&A...453..635M}, the warp is more prominent at larger
distances from the Sun (beyond a few kpc), plus heavy {\em local}
extinction may prevent any such signature showing in the bright
emitter population at some longitudes -- particularly those well
inside the Solar Circle ($\ell \ltappeq 60^{\rm o}$). For example, the
heavy local ($\ltappeq 2$~kpc) extinction associated with the Aquila
Rift may be responsible for the deficit of bright emitters at positive
latitudes around $\ell \simeq 30^{\rm o}$ (cf. Figures~7 and 8 in
\citealt{2006A&A...453..635M}).

\section{The Photometric Properties of H$\alpha$ Emitters}

\subsection{The $(r^\prime - H\alpha)$ vs $(r^\prime - i^\prime)$
  Distribution of H$\alpha$\ Emitters}

\begin{figure}

\centering

\includegraphics[angle=0,width=1.\linewidth]{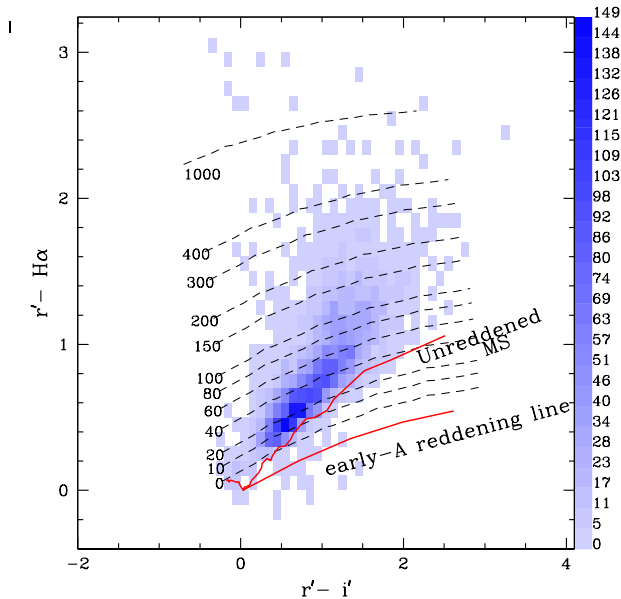}

\caption{The distribution of H$\alpha$ emitters in the 
($r^\prime-\rmn{H}\alpha$) versus ($r^\prime-i^\prime$) 
plane. The shade of blue assigned to each grid square in the
colour-colour plane indicates the number of emitters it contains, as
defined by the contrast scale on the right. The expected locations of
the unreddened main sequence and the early-A star reddening line are
also shown. The dashed lines represent lines of constant
H$\alpha$ equivalent width (in \AA\, as labelled) for early-type
stars. These lines move from left to right with increasing
reddening. All emitters are counted here, including those flagged with
'c' in Table~1.}

\label{fig:sel_plane}

\end{figure}

Fig.\,\ref{fig:sel_plane} shows the distribution of catalogue objects
in the $(r^\prime - H\alpha)$ vs $(r^\prime - i^\prime)$ colour-colour
plane. This is the plane used by our selection algorithm (cf Fig.\,
\ref{fig_selection}). We recall that our algorithm works by
identifying objects located above the upper stellar locus in each
field, and that this locus is usually composed of unreddened
main-sequence stars (Section~\ref{select}). In line with this,
Fig.\,\ref{fig:sel_plane} shows that the vast majority of our
H$\alpha$-excess sources lie above the expected location of
the unreddened main sequence. The few objects below this line 
may still be genuine emitters (e.g. they could come from fields in
which the absolute photometry is somewhat compromised), but their
status is less certain.

Following \citet{2006MNRAS.369..581W} and \citet{2005MNRAS.362..753D},
we have also indicated in Fig.\,\ref{fig:sel_plane} lines of constant
H$\alpha$ EW for stars with relatively blue spectral energy
distributions (see \citealt{2005MNRAS.362..753D} for details). This
shows that a reddened early-type emission line star must have
considerably higher EW to make our selection cut than an unreddened
one. 

\subsection{The Magnitude Distribution of H$\alpha$ Emitters}

\begin{figure}

\centering

\includegraphics[angle=0,width=1.0\linewidth]{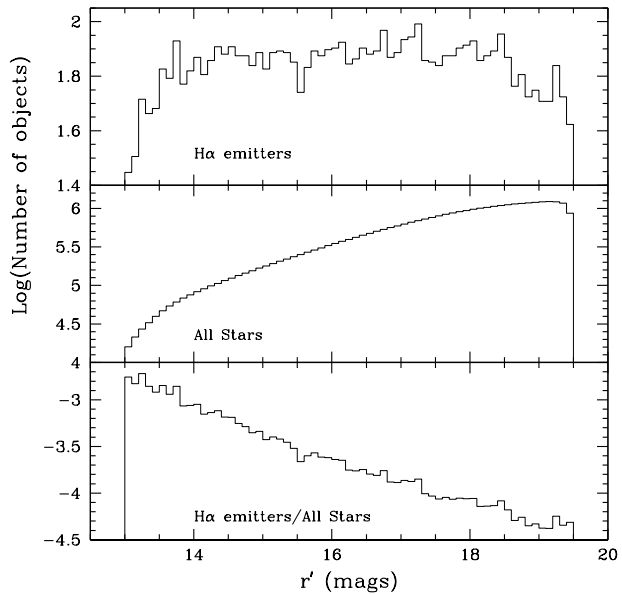}


\caption{The $r^\prime$ band magnitude distribution of H$\alpha$
  emitters in the catalogue along with the distribution of the
  all-star sample and the ratio of the two.}

\label{fig:mag_dist}

\end{figure}

In the top panel of Fig.\,\ref{fig:mag_dist}, we show the $r^\prime$
magnitude distribution of our catalogue objects. The fact that this
distribution is flat in the range $14 \ltappeq
r^\prime \ltappeq 18$\ is not an indication of completeness. After
all, as shown in the middle panel of Fig.\,\ref{fig:mag_dist}, the
total number of stars (not just emitters) rises steeply towards
fainter magnitudes almost all the
way to the catalogue limit. Thus the fraction of objects 
selected as H$\alpha$\ emitters is a strongly decreasing function of
magnitude (bottom panel of Fig.\,\ref{fig:mag_dist}). This mainly
reflects the fact that it is much easier to select bright H$\alpha$
emitters than faint ones.

This statement can be quantified. The minimum H$\alpha$ EWs for
inclusion in the catalogue can be estimated from the IPHAS photometry
using the techniques described in \citet{2006MNRAS.369..581W} and
\citet{2005MNRAS.362..753D}; see also Fig.\,\ref{fig:sel_plane}). In
principle, this limiting EW depends on both the intrinsic SED shape
and the observed broad-band colour. Here, we simply adopt $(r^\prime -
i^\prime) \simeq 1$ as representative for our catalogue objects and
take their SEDs to be reddened Rayleigh-Jeans spectra (many
of them will be early-type emission line stars; see
Section~\ref{spec}). {\em For these choices of SED and colour}, we
then estimate limiting EWs of 
$\sim15$\AA~ in the range $r^\prime = 13-17.5$,  $\sim25$\AA\ in the
range $r^\prime = 17.5-18.5$, and $\sim50$\AA\ in the $r^\prime =
18.5-19.5$. 

\subsection{The $r^\prime-\rmn{H}\alpha$ Distribution of H$\alpha$\ Emitters}

\begin{figure}

\centering

\includegraphics[angle=0,width=1.0\linewidth]{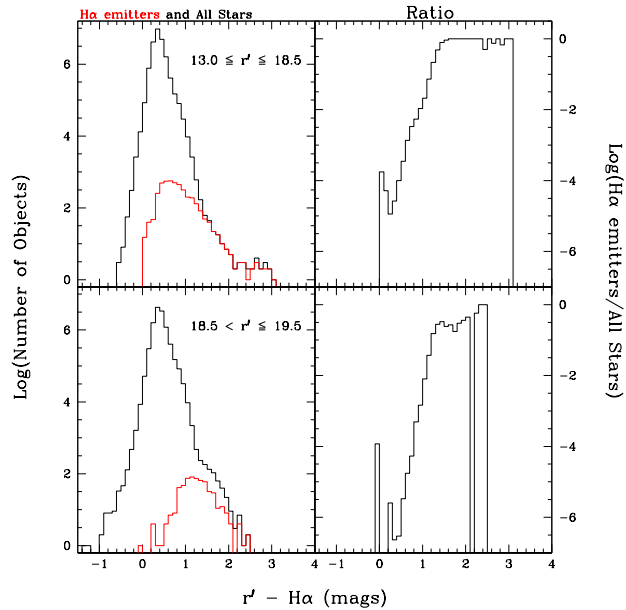}


\caption{The ($r^\prime-\rmn{H}\alpha$) distribution of H$\alpha$
  emitters in the catalogue along with the distribution of the
  all-star sample and the ratio of the two. The left
  hand panels show the distribution of the H$\alpha$ emitters (red
  histograms), overplotted on top of the distribution of the
  corresponding all-star samples (black histograms). The right hand
  panels show the ratio of the emitter and all-star distributions.
  The top panels show the distributions of bright objects ($r^\prime
  \leq 18.5$; the bottom panels show the distributions for 
  fainter objects.}

\label{fig:rminh_dist_bf}

\end{figure}

Fig.\,\ref{fig:rminh_dist_bf} shows the ($r^\prime-\rmn{H}\alpha$)
distribution of catalogue objects, separated into two coarse magnitude
bins. The emission line star populations produce a distinct
``shoulder'' in the ($r^\prime-\rmn{H}\alpha$) distributions of the
all-star samples. The fraction of stars selected as emitter rises 
monotonically with H$\alpha$\ excess, until at sufficiently large
excesses, almost all objects are selected. For bright stars, this
happens beyond $(r^\prime-\rmn{H}\alpha) \simeq 1.6$; for faint stars,
it only happens beyond $(r^\prime-\rmn{H}\alpha) \simeq 2.3$. More
generally, at any given H$\alpha$\ excess, the fraction of objects
selected is larger for bright objects than for faint ones. The peak in
the number distribution of bright emitters also lies at substantially
smaller ($r^\prime-\rmn{H}\alpha$) values. 

All of these results are in line with expectations: once the
H$\alpha$-excess of emission line objects becomes significantly larger 
than the dispersion of ordinary stars in ($r^\prime-\rmn{H}\alpha$),
the probability that they will be selected rises monotonically with
increasing excess. At any given H$\alpha$\ excess, this probability is
higher for bright objects, since the dispersion caused by photometric
scatter is smaller for them. On the other hand, the {\em intrinsic}
H$\alpha$-excess distribution of Galactic emission line stars is likely to 
rise monotonically towards smaller excesses (i.e. there are more weak 
line emitters than strong ones). The location of the peaks in the
observed ($r^\prime-\rmn{H}\alpha$) distributions of emitters is thus
set by the competition between these effects.

\subsection{The $r^\prime-i^\prime$ Distribution of H$\alpha$\ Emitters}
\label{bimodal}

\begin{figure}

\centering

\includegraphics[angle=0,width=1.0\linewidth]{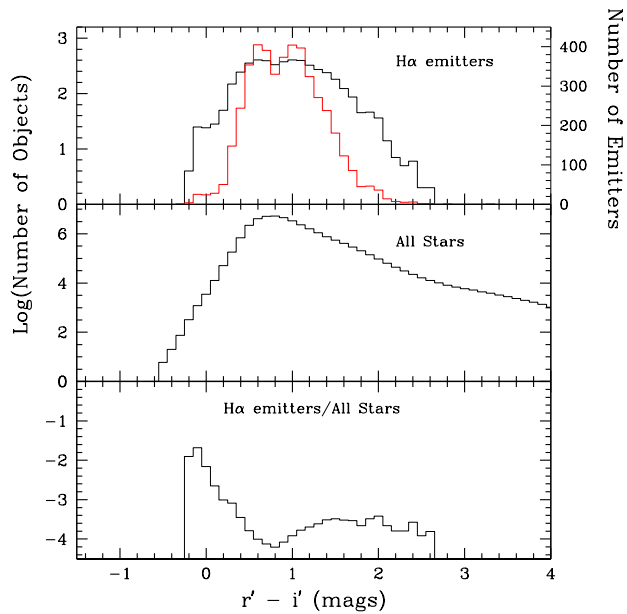}


\caption{The ($r^\prime-i^\prime$) distribution of H$\alpha$ emitters
  along with the distribution of the all-star sample and the ratio of
  the two. The black distributions are shown on a logarithmic scale
  (left axes); the red distribution in the top panel shows the
  emitter distribution on a linear scale (right axis). }

\label{fig:rmini_dist}

\end{figure}

Fig.\,\ref{fig:rmini_dist} shows the ($r^\prime-i^\prime$)
distribution of H$\alpha$ emitters.  The top panel shows the
distribution of the catalogue objects themselves, the middle panel
shows the distribution of the all-star sample, and the bottom panel
shows the ratio of the two distributions. 

The emitter distributions are clearly bimodal, as is particularly
obvious for their fractional incidence relative to other stars. The
blue peak is almost certainly dominated by early-type emission line
stars (see Section~\ref{spec}). The existence of the red peak may
reflect an increasing contribution of other objects, such as young/active
late-type stars to the emitter population. However,  
spectroscopic follow-up of a significant sample of faint, red
catalogue objects will be required to determine the make-up of this
population with confidence.

\section{Discussion}

\subsection{The Nature of H$\alpha$-Excess Objects}

\label{nature}

One of the most important questions regarding our emitter catalogue
concerns the nature and fractional contribution of 
different stellar populations to the overall sample. Below, we will
attempt a preliminary answer to this question based on matches
to existing star catalogues and initial spectroscopic follow-up
observations of our own.

\subsubsection{Matches with the KW99 Catalogue}

\label{kw99_match}

Twenty-seven percent of KW99 objects are fainter than 13th\mags, and
we expect a significant fraction of these to be present in our
catalogue also. We have therefore performed a 
positional cross-match between objects in the two catalogues. 
Magnitude information was not considered when matching, and only
those 3543 objects in KW99 that fall inside the area covered by our
catalogue were used.  

In order to select an appropriate matching radius, and get a handle on
the number of spurious matches, we created a 
mock catalogue in which all positions in the IPHAS H$\alpha$\ emitter
catalogue were shifted by $0.5^\circ$\ in Galactic latitude and longitude. 
We found that the matching radius that maximizes the difference
between matches to the real data and matches to the mock catalogue is
7.4\,arcmin. This is about 5 
times the positional uncertainty suggested by KW99 for stars belonging 
to their lowest accuracy category. With this matching radius, 1056 
IPHAS emitters are matched to KW99 objects, whereas 536 matches are 
found for the mock catalogue. We therefore expect that $\sim$
500 objects ($\sim$\ 10 per cent of the IPHAS sample) are common to
both catalogues. Since a 50 per cent false match probability is not
very helpful when it comes to identifying specific objects in common, 
we only flag matches to KW99 in our catalogue if the coordinates agree
to better than 30 arcsec. With this matching radius, we expect only
about 3 false matches amongst the 252 flagged objects (which represent
about 5 per cent of the IPHAS-based catalogue).  

\begin{figure}

\centering

\includegraphics[angle=0,width=1.0\linewidth]{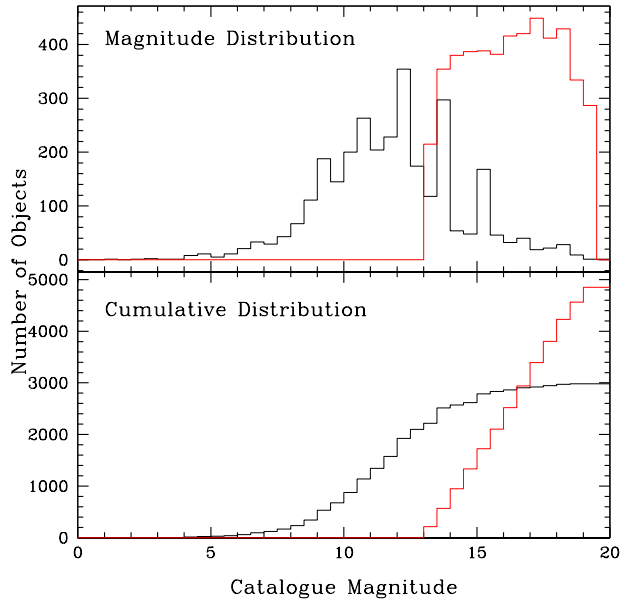}

\caption{The magnitude distribution of the IPHAS catalogue of
  H$\alpha$ emitters (red line) compared to the magnitude distribution
  of objects in the KW99 catalogue (black line). The IPHAS magnitudes
  shown correspond to the $r^\prime$ band, and the majority of the KW99
  magnitudes correspond roughly to V band (based on either photoelectric,
  photographic or photovisual measurements). The top panel shows the
  magnitude distribution, and the bottom panel shows the cumulative
  distribution. All IPHAS emitters are shown here, including those
  flagged with 'c' in Table~1.}

\label{fig:kohoutek_match}

\end{figure}

It is interesting to compare the expected number of objects common to
both catalogues ($\sim500$) to the magnitude distributions of the two
catalogues. These are shown in Fig.\,\ref{fig:kohoutek_match} in both
differential (top panel) and cumulative form (bottom panel). In order 
to allow for a meaningful comparison, we have only considered the 2984
KW99 objects that fall inside the spatial area of the IPHAS catalogue
and also have magnitude estimates. Ignoring the differences between
the IPHAS and KW99 
photometric band-passes for the moment, the magnitude distributions
would suggest that $\sim$ 900 objects should fall within the area
and magnitude range covered by both catalogues. Allowing for the
difficulty in comparing source magnitudes 
between the two catalogues, and the fact that $\sim400$ of the
overlapping objects in the KW99 magnitude distribution are located
very close to the bright cut-off of the IPHAS catalogue, we suspect
that this number is consistent with the approximate value derived from
the positional cross-matches.  

\subsubsection{Matches with SIMBAD}

\label{simb_match}

We have also cross-matched the objects in our H$\alpha$-excess
catalogue against previously known sources in SIMBAD. As a starting
point, we initially 
searched for all objects in SIMBAD within 1\,arcmin of the IPHAS
positions. The results of this initial search showed that the majority
of matches fall within 8 main categories. In order of most matches,
these categories are: emission line stars (Em*), infrared sources
(IR), normal stars (*), stars in clusters (*iC), stars in associations
(*iA), radio sources (Rad), X-ray sources (X) and variable stars
(V*). 

As in Section~\ref{kw99_match}, we used mock data sets to determine
appropriate matching radii and estimate the number of spurious
matches. In this case, the matching radius for each of the 8 main
categories was chosen 
to keep the fraction of spurious matches below about 5 per cent. We
allowed two exceptions to this 
rule: (i) the matching radius to X-ray sources was increased to allow
an expected 10 per cent contamination fraction; (ii) the matching to
IR sources was increased to allow an expected 8 per cent contamination
fraction. In both cases, this was done because the total number of
matches was still increasing significantly as a function of matching
radius at the 5 per cent contamination level. 

The resulting matching radii for each object class and the number of
matches obtained are shown in the top part of
Table~\ref{main_match}. Within each object class, we have also looked
for objects listed as emission line objects. This was done by checking
the ``other object type'' field in SIMBAD. 

As expected, the highest number of matches belong to the Em*
category. The majority of these objects come from KW99, although there 
are a few additional objects from other sources. We have also included
in Table~\ref{main_match} (in the row that is directly above the first
set of totals) the number of matches to objects whose main object type
is not Em*, but which have Em* defined as a secondary object type in
SIMBAD.  These are labelled as Em*2, and the matching radius used for
these objects was the same as the Em* objects. A total of 316 IPHAS
emitters have been matched with Em*/Em*2 in SIMBAD. This 7 per cent
matching fraction is broadly consistent with the results of the
previous section (a 5-10 per cent matching fraction depending on
matching radius). Thus in the interval between the publication of KW99
and the present catalogue, only a small number of additional emitters
appear to have been found in the Northern Milky Way.   

Perhaps the most interesting matches are those to Infra-Red (IR),
X-ray and Radio sources. The majority of the IR sources are IRAS
identifications. These sources are often young stars still shrouded in
dust, and the sources matched with IPHAS H$\alpha$ emitters are likely T
Tauri or Herbig-type objects. The 20 arcsec matching radius used for
the IR sources is comparable to the positional uncertainty of IRAS
sources (typically between 2 and 16 arcsec, but up to 1 arcmin in
extreme cases). The X-ray matches are all ROSAT sources, and the 20
arcsec matching radius used is similar 
to the typical positional uncertainties of ROSAT sources
\citep{1999A&A...349..389V}. Note that 14 of the 26 X-ray sources, the
majority of the IR sources and all of the radio sources have no other
object associations or source classifications in SIMBAD. These sources
are excellent follow-up candidates. Indeed, one of the X-ray sources has
already been identified as a new CV candidate (IPHAS
J052832.69+283837.6; Witham et al, submitted to MNRAS).  

\begin{table*}

\begin{minipage}{\linewidth}

\begin{center}

\caption{\label{main_match} A summary of the results obtained from a
  positional cross-match between the catalogue of IPHAS emitters and
  SIMBAD. SIMBAD categories of objects are listed in the first
  column. The main categories of SIMBAD objects are listed in the top
  part of the table. The objects not included in the 8 main object
  types are listed in bottom part of the table. The SIMBAD short-hand
  object classifications are given in the second column. The final
  column lists the SIMBAD matches which have emission-line star listed
  as either the main object type or secondary object type.} 

\begin{tabular}{llllll}

\hline Main object categories & associated SIMBAD object types\tablenotemark{a} & matching & \# of IPHAS & \# of SIMBAD    & \# SIMBAD\\

                              &                                                 & radius   & emitters with & objects matched & objects \\

                              &                                                 & (arcsec) & SIMBAD      & to IPHAS        & classed as \\

                              &                                                 &          & matches     & emitters        & emitters\\

\hline

Emission-line Star & EM* & 45 & 288 & 360 & 360\\

Infra-Red source & IR & 20 & 106 & 111 & 1\\

Star & $^\ast$ & 3 & 22 & 22 & 0\\

Star in Cluster & $^\ast$iC & 5 & 21 & 22 & 0\\

Star in Association & $^\ast$iA & 3 & 15 & 16 & 0\\

Radio-source & Rad & 15 & 7 & 9 & 0\\

X-ray source & X & 20 & 26 & 26 & 0\\

Variable Star & V* & 20 & 13 & 14 & 4\\

Emission-line Star\tablenotemark{b} & Em*2 & 45 & 28 & 35 & 35\\

\hline

Totals & & & 526 & 615 & 400\\

\hline

\\

\\ \hline

Additional Object Categories\\

\hline

Nebulae &  {\tiny{BNe, HH, HII, PN, PN?, RNe, Cld, EmO}} & 10 & 21 & 24 & 4\\

Interacting Binaries &  {\tiny{CV*, DQ*, DN*, Al*, No*, NL* Sy*, XB}} & 10 & 33 & 34 & 5\\

Young Stars &  {\tiny{Y*O, TT*, FU*, Or*}} & 10 & 16 & 16 & 1\\

Other Variable Star &  {\tiny{Pu*, RI*}} & 10 & 5 & 5 & 2\\

Other Star &  {\tiny{Be*, C*, $^\ast$i*, WD*, WR*}} & 10 & 10 & 11 & 3\\

Misc & {\tiny{Cl*, G, Mas, ?, UV}} & 10 & 11 & 14 & 0\\

\hline

Totals & & & 96 & 104 & 15\\

\hline

\footnotetext[1]{See http://vizier.u-strasbg.fr/viz-bin/Otype?X for
a description of the SIMBAD classification scheme.}

\footnotetext[2]{These emission-line stars are those objects whose
  main object type is not listed as emission-line star but have
  emission-line star listed as a secondary object type.} 

\end{tabular}

\end{center}

\end{minipage}

\end{table*}

There are also a significant number of matches to SIMBAD objects that
do not belong to the 8 main object classes listed above. Due to the
small number of matches for each of these object types, it has not
been possible to establish the best matching radius for each type. We
have therefore simply adopted a 10\,arcsec matching radius and
searched for any IPHAS objects matched with SIMBAD objects other than
those in the classes above. 

Since rather a lot of different object types have been uncovered by
this exercise, we only present a concise summary in the bottom part of
Table~\ref{main_match}. Thus we have grouped the various SIMBAD types  
into a coarser classification scheme that combines related classes
under one heading. We find that interacting binaries are the dominant
population amongst these other matches. This is not surprising given
that the majority of these objects are known to be H$\alpha$
emitters. Several nebulae are also detected, which reinforces the fact
that the catalogue is not completely free of extended
objects. There are also matches to young stars and variable stars, as
well as to variety of miscellaneous objects, including white dwarfs,
carbon stars, Wolf-Rayet stars, masers, etc. 

We finally note that there are only 4 matches to objects labelled as
Be stars in SIMBAD, even though our own follow-up suggests that many
objects in our catalogue belong to this class (see Section~\ref{spec}
below). The explanation is probably that most of the bright Be stars
are simply labelled as Em* in SIMBAD, whereas most of the faint
objects remain to be identified. 

In total, we have found SIMBAD matches to 519 IPHAS objects. (This
is lower than the sum of the total values in Table~\ref{main_match},
because some IPHAS sources are matched to multiple SIMBAD objects with
different classification.) This total number of matches suggests that
$\simeq$\ 90 per cent of the IPHAS emitters are previously unknown
systems. Thus IPHAS is uncovering a large new population of H$\alpha$
emission line stars.

\subsubsection{Spectroscopic Follow-up}
\label{spec}

Given that only a small fraction of the catalogue consists of
previously known objects, a large-scale spectroscopic follow-up effort
is necessary to determine the true make-up of the catalogue. Such an
effort is underway, using both long-slit and multi-fibre
spectroscopy. Here, we merely note that a first analysis of spectra 
for $\sim300$ catalogue objects brighter than $r^\prime=18$\ 
confirms $\simeq97$\ per cent of them as H$\alpha$\ emitters, with 
$\sim85$ per cent belonging to broad class of early-type emission
line stars \citep{witham07}. This dominance of early-type emitters is
consistent with the results of KW99. Interacting binaries, late-type
stars and young stellar objects make up smaller fractions of this
sample.

However, it is important to keep in mind that this early spectroscopic
sample is limited to bright emitters and, partly because of this, also 
biased towards blue objects \citep{witham07}. Thus the make-up of the
fainter and redder H$\alpha$-excess population still needs to be
explored. Broadly speaking, we should still expect to find many
early-type emission line stars as we push to fainter
magnitudes, because both the closest and the more distant spiral arms
are expected to harbour significant populations of these
objects. However, at these fainter magnitudes, we may also expect to
find an increasing proportion of interacting binaries, late-type
stars, and T Tauri stars in star-forming regions. The often close
grouping of the faint emitters in the 2-D distribution of emitters in
the Galactic Plane shown in Fig.\,\ref{fig:gal_4_panel} provides
evidence for the increased presence of emitters in star forming
regions and associations at faint magnitudes. 

\section{Conclusions}

\label{conc}

A catalogue of H$\alpha$ emitters selected from existing IPHAS
photometry has been presented and analysed.  The catalogue contains
4853 objects in the magnitude range $13 \leq r^\prime \leq
19.5$\ selected from $\sim$ 150 million objects in total. The
average surface density of these H$\alpha$-excess objects is $\sim3$
emitters per square degree. The true surface density of
H$\alpha$-excess objects in the Galactic Plane is expected to be
higher than this, due to the bias in the selection technique against
detecting reddened earlier-type and/or weak-lined emitters, especially
in regions where the surface density of emitters is high. The present
catalogue is compiled from $\simeq$\ 80 per cent of the final IPHAS
survey area, leading to an expected number of at least $\sim6000$
H$\alpha$ emitters from 
IPHAS, even with the very conservative selection procedure adopted
here. Preliminary spectroscopic follow-up observations suggest that,
at least to $r^\prime \simeq 18$, more than 95\% of our catalogue
objects are genuine H$\alpha$ emitters.

The ($r^\prime-i^\prime$) distribution of H$\alpha$ emitters is
bimodal, with a minimum near the peak in the distribution of the
all-star sample (non-emitters). The blue peak in the emitter
distribution is dominated by early-type emission line
stars. The red peak is probably due to an increasing contribution of
other types, most notably young/active late-type stars. The spatial
distribution of catalogue objects shows tentative evidence for the
Galactic warp, and also exhibits overdensities towards OB
associations, clusters, bright rim clouds and the spiral arms of the
Galaxy.

The catalogue has been cross-matched with the emission line star
catalogue of \citet{1999A&AS..134..255K} and also with SIMBAD. We find
that a maximum of $\simeq$ 10 per cent of the IPHAS H$\alpha$\ emitters
are previously known objects. The SIMBAD matches include several
previously known interacting binaries, nebulae and young stellar
objects. Many matches are to objects whose detailed classification is
unknown, including a sample of infrared, radio and X-ray sources.   

\section*{Acknowledgments}

Based in part on observations made at the Isaac Newton Telescope, which
is operated on the island of La Palma by the Isaac Newton Group in the
Spanish Observatorio del Roque de los Muchachos of the IAC. This
paper makes use of data obtained at the FLWO Observatory, a facility
of the Smithsonian Institution. ARW was supported by a PPARC
Studentship. BTG was supported by a PPARC Advanced Fellowship.  DS
acknowledges a Smithsonian Astrophysical Observatory Clay Fellowship
and a PPARC/STFC Advanced Fellowship. PJG is supported by NWO VIDI
grant 639.042.201. This research has made use of the SIMBAD database,
operated at CDS, Strasbourg, France.  This research has made use of
NASA's Astrophysics Data System.

\bibliographystyle{mn_new}

\bibliography{mn-jour,arwbib}

\bsp

\label{lastpage}

\end{document}